\documentstyle[12pt]{article}
\newcommand{\bc}{\begin{center}}
\newcommand{\ec}{\end{center}}
\begin{document}
{}\hfill   hep-th/9903083

\bigskip

\bc {\Large\bf Symmetries and causes of the coincidence of the radiation spectra
of mirrors and charges in~1~+~1~and~3~+~1~spaces}
\footnote{Published in Zh.Eksp.Teor.Fiz. 114, 46 (1998)}
\ec

\bc {\large   V.I.Ritus}\footnote{E-mail: ritus@lpi.ac.ru}
\ec

\bc {\it Lebedev Physical Institute, 117924 Moscow, Russia} \ec

\bc              {\normalsize   Abstract} \ec

This paper discusses the symmetry of the wave field that lies to the
right and left of a two-sided accelerated mirror in 1 + 1 space and
satisfies a single condition on it.  The symmetry is accumulated in
the Bogolyubov matrix coefficients $\alpha$ and $\beta$ that connect
the two complete sets of solutions of the wave equations.  The
amplitudes of the quantum processes in the right and left half-spaces
 are expressed in terms of $\alpha$ and $\beta$ and are related to
 each other by transformation (12).  Coefficient
$\beta_{\omega'\omega}^*$ plays the role of the source amplitude of a
pair of particles that are directed to opposite sides with
frequencies $\omega$ and $\omega'$ but that are in either the left or
the right half-space as a consequence of the reflection of one of
them.  Such an interpretation makes $\beta_{\omega'\omega}^*$
observable and explains the equalities, given by Eq.~(1) and found
earlier by Nikishov and author [Zh. Eksp. Teor. Fiz. 108, 1121
(1995)] and by author [Zh. Eksp. Teor. Fiz. 110, 526 (1996)]
that the radiation spectra of a mirror in 1+1 space coincide with
those of charges in 3 + 1 space by the fact that the moment of the
pair emitted by the mirror coincide with the spin of the single
particle emitted by the charge.

\section{Introduction}
It was found in Refs.[1] and [2] that the spectra of bosons and fermions emitted
by an accelerated mirror in 1 + 1 space coincide with the spectra of photons and
scalar quanta emitted by electric and scalar charges in 3 + 1 space when the
latter move along the same trajectory as does the mirror.
Namely, the Bogolyubov coefficients $\beta_{\omega'\omega}^{B,F}$ that
describe the spectra of the Bose and Fermi radiations of an accelerated mirror
and the Fourier transforms of the density of the 4-current $j_\alpha(k_+,k_-)$
and the scalar charge density $\rho(k_+,k_-)$ that describe the spectra of the
photons and scalar quanta emitted by electric and scalar charges are connected by
the relationships
\begin{equation}
\vert\beta^B_{\omega'\omega}\vert^2=\frac1{e^2}\vert j_\alpha(k_+,k_-)\vert^2,
\qquad \vert\beta^F_{\omega'\omega}\vert^2=\frac1{e^2}\vert\rho
(k_+,k_-)\vert^2.
\end{equation}
It is assumed here that the components $k_\pm=k^0\pm k^1$ of the wave 4-vector
$k^\alpha$ of the quantum emitted by the charge are identified with the doubled frequencies $\omega$ and $\omega'$
of the quanta emitted by the mirror:
\begin{equation}
2\omega=k_+,\qquad 2\omega'=k_-,
\end{equation}
and $e$ is the electrical or scalar charge in Heaviside units.

However, there is a substantial physical difference between the right-hand and
left-hand quantities in Eqs.~(1), i.e., between the radiation spectra of the
charges and of the mirror. Whereas the former are the distribution of the mean
number of radiated quanta over the two independent components $k_+$ and $k_-$
of the wave vector of the quantum (as a consequence of the azimuthal symmetry
of the radiation, there is no dependence on the third independent variable),
the latter have a more complex interpretation.  Actually, they will be the
spectra of the mean number of quanta emitted by the mirror to the right only
after integration over frequency $\omega'$ [3]:
\begin{equation}
d\bar n_\omega=\frac{d\omega}{2\pi}\int\limits_0^\infty \frac{d\omega'}{2\pi}
\vert\beta_{\omega'\omega}\vert^2.                         
\end{equation}

If the mirror is two-sided and infinitely thin, then, besides the quanta emitted to
the right with the spectrum given by Eq. (3), it will (as we shall see) also
emit quanta to the left with the spectrum
\begin{equation}
d\bar n'_{\omega'}=\frac{d\omega'}{2\pi}\int\limits_0^\infty
\frac{d\omega}{2\pi} \vert\beta_{\omega'\omega}\vert^2.          
\end{equation}

The thought naturally arises whether it is not possible to regard the quantity
\begin{equation}
\vert\beta_{\omega'\omega}\vert^2 \frac{d\omega d\omega'}{(2\pi)^2}
\end{equation}
as the mean number of pairs of quanta, one of which, with frequency $\omega$
in the interval $d\omega$, is emitted by the mirror to the right, while the other,
with frequency $\omega'$ in the interval $d\omega'$, is emitted to the left.  In
this case, two frequencies $\omega$ and $\omega'$ would be observable,
characterizing one event: the emission of a pair of quanta by the mirror, in the
same way as two components $k_+$ and $k_-$ also characterize one event: the
emission of one quantum by a charge.  As we can see, with certain nontrivial
complications, such a treatment is actually valid.  In any case, the mirror emits
quanta in pairs.

It is elucidated that this circumstance helps to understand another
difference between the coincident spectra of a charge and a mirror. While
the bosons and fermions emitted by a mirror have a spin of 0 and 1/2, the
photons and scalar quanta emitted by electrical and scalar charges
have a spin of 1 and 0.  Even though the quanta have different spin, the
radiation spectra of the charges coincide with the boson and fermion spectra of
the mirror.

This coincidence is explained by the fact that, unlike charges, the mirror emits
particles in pairs, and a pair of spinless bosons can have a total moment of 1,
while a pair of fermions can have a total moment of 0.  Then the moment of the
pair emitted by the mirror coincides with the spin of the particle emitted by the
charge.  The fact that, upon reflection, $\beta_{\omega'\omega}^B$ behaves like
a pseudoscalar while $\beta_{\omega'\omega}^F$ behaves like a scalar can serve
as an indirect confirmation of this (see Sections 2 and 4).

It is shown in Section 2 that the system of Bogolyubov coefficients obtained
for a right-sided mirror (i.e., for the field to the right of a mirror with a
boundary condition on it), because of the properties of mirror symmetry, also
describes the processes in the field to the left of a mirror with the same
boundary condition.  In other words, the same system of Bogolyubov coefficients
characterizes the behavior of the field in all of space---both to the right and
to the left of a two-sided mirror.

Section 3 describes the connection between the integral quantities that
characterize the radiation of a two-sided mirror, their behavior under certain
space--time transformations, and the symmetry (or asymmetry) of the
 space--time regions where they are formed.

The symmetry of the Bogolyubov coefficients reflects the symmetry of two
inequivalent complete systems of solutions of wave equation, definite and smooth
in all of 1 + 1 space, satisfying inside it---on the trajectory of the mirror---a
single condition and characterized by propagation of a monochromatic
component of each solution toward the right in one system and toward the left in
the other.  When the field is quantized and when the usual comparison of
monochromatic plane waves to particles is made, these two systems of solutions
form {\it in} and {\it out} systems for the field to the right of the trajectory and
{\it out} and {\it in} systems for the field to the left of it.  Therefore, the
quantum processes in the field to the right and to the left of the mirror are
independent, even though they are described by a single system of Bogolyubov
coefficients.  In particular, the particle-production amplitudes to the right and to
the left of the mirror, the single-particle scattering amplitudes in these
regions, etc.\ are connected with transformation~(12).  Such amplitudes,
certain frequency distributions, and also the distribution of pair-production
probabilities over the number of pairs, which is invariant relative to
transformation (12), are computed in Section 4.  It is shown that
$\beta_{\omega'\omega}^*$ plays the role of the source amplitude of a pair of
particles potentially emitted to the right and to the left with frequencies
$\omega$ and $\omega'$, with the spin of a boson pair equaling 1, while that of
the fermion pair equals 0.

In the last section, Sec. 5, a similar method is used to treat the emission by
an accelerated mirror of pairs the particle and antiparticle of which are not
identical.

A system of units in which $\hbar=c=1$ is used in this article.  To simplify the
formulas in Sections 4 and 5, the frequencies are considered discrete,
integration over $d\omega/2\pi$ is replaced by summation over $\omega$, and
the delta function $2\pi\delta(\omega-\omega'')$ is replaced by the Kronecker
symbol $\delta_{\omega\omega''}$.

\section{Symmetry of the Bogolyubov coefficients and radiation of
accelerated two-sided mirror}
Let us consider the connection between radiation spectra and other quantities
in two problems in which the mirror trajectories $x=\xi_1(t)$ and $x=\xi_2(t)$
differ by reflection: $\xi_1(t)=-\xi_2(t)$. Then, if the first trajectory is
described on the plane of variables $u=t-x$, $v=t+x$ by the function $v=v_1=f(u)$,
the second trajectory will be described by the function inverse to it
$v=v_2=g(u)$,  $g\mbox{\boldmath$($}f(u)\mbox{\boldmath$)$}=u$.

The Bogolyubov coefficients, defined as in Refs.~[1] and [2] for the field to
the right of the mirror [see also Eqs.~(33) and (34)],
\begin{equation}
\alpha^B_{\omega'\omega}[f],\quad \beta^{B*}_{\omega'\omega}[f]=
\pm\sqrt{\frac{\omega}{\omega'}}\int_{-\infty}^\infty du\exp\,[\mp i\omega u+
i\omega' f(u)]=                                                     
\end{equation}
\begin{equation}
\hspace{2,2cm}=\sqrt{\frac{\omega'}{\omega}}\int_{-\infty}^\infty dv\exp\,[
i\omega' v\mp \omega g(v)],                                         
\end{equation}
being functionals of the trajectory, when $f(u)$ is replaced by
$g(u)$ and consequently $g(v)$ is replaced by $f(v)$, transform to
\begin{equation}
\alpha^B_{\omega'\omega}[g]=\alpha^{B*}_{\omega\omega'}[f],\quad
\beta^B_{\omega'\omega}[g]=-\beta^B_{\omega\omega'}[f].              
\end{equation}
Similarly, the Bogolyubov coefficients for a fermion field
\begin{equation}
\alpha^F_{\omega'\omega}[f],\quad \beta^{F*}_{\omega'\omega}[f]=\int_{-\infty}
^\infty du\sqrt{f'(u)}\exp\,[\mp i\omega u+i\omega' f(u)]=            
\end{equation}
\begin{equation}
\hspace{2,2cm}=\int_{-\infty}^\infty dv\sqrt{g'(v)}\exp\,[i\omega' v\mp i\omega
g(v)],                                                                 
\end{equation}
when the trajectory is replaced by its mirror reflection, transform to
\begin{equation}
\alpha^F_{\omega'\omega}[g]=\alpha^{F*}_{\omega\omega'}[f],\quad
\beta^F_{\omega'\omega}[g]=\beta^F_{\omega\omega'}[f].                 
\end{equation}
The matrix notations for the Bogolyubov coefficients make it possible to write
the transformations of Eqs.~(8) and (11), i..e, the transition from trajectory
$f(u)$ to $g(u)$, in the form
\begin{equation}
\alpha\to\alpha^+,\qquad \beta\to\mp\tilde \beta,                      
\end{equation}
where the upper and lower signs here and subsequently correspond to Bose and
Fermi fields.

At the same time, according to the expansions given by Eqs.~(33) and (34),
$\alpha_{\omega'\omega}$ and $\beta_{\omega'\omega}$ are the amplitudes of
the waves with frequencies $\omega'$ and $-\omega'$ contained in the incident
part of the {\it out\/} wave with frequency $\omega$, while
$\alpha_{\omega'\omega}^*$ and $\mp\beta_{\omega'\omega}$ are the
amplitudes of the waves with frequencies $\omega$ and $-\omega$ contained in
the reflected part of the {\it in\/} wave with frequency $\omega'$.  Therefore,
amplitudes $\alpha_{\omega'\omega}^*[g]$ and $\mp\beta_{\omega'\omega}[g]$
describe the generation by a right-sided mirror on trajectory $g(u)$ of waves
escaping to the right with frequencies $\omega$ and $-\omega$ when wave with
frequency $\omega'$ incident from right to left is absorbed.  From purely
geometrical considerations, they must coincide with the amplitudes for the
mirror-symmetric process---the generation by a left-sided mirror on trajectory
$f(u)$ of waves escaping to the left with frequencies $\omega$ and $-\omega$
when a wave incident from left to right with frequency $\omega'$ is absorbed.
Then, according to Eqs.~(8) and (11), these last are also equal to
$\alpha_{\omega\omega'}[f]$ and $\beta_{\omega\omega'}[f]$ or equal to
$\alpha_{\omega'\omega}[f]$ and $\beta_{\omega'\omega}[f]$ if the frequencies
of the monochromatic waves propagating to the right and to the left are denoted
as $\omega$ and $\omega'$, as was assumed for the field to the right of the
trajectory.

Thus, for the field to the left of a mirror moving along trajectory $f(u)$,
$\alpha_{\omega'\omega}[f]$ and $\beta_{\omega'\omega}[f]$ are the amplitudes
of waves with frequencies $\omega'$ and $-\omega'$ contained in the reflected
part of the {\it in} wave with frequency $\omega$, while
$\alpha_{\omega'\omega}^*[f]$ and $\mp\beta_{\omega'\omega}[f]$ are the
amplitudes of waves with frequencies $\omega$ and $-\omega$ contained in the
incident part of the {\it out\/} wave with frequency $\omega'$.  Therefore, the
matrix that connects the {\it in\/} and {\it out\/} waves of the field to the left of
the mirror differs from the analogous matrix for the field to the right of it by
transformation~(12).

So the transition from trajectory $f(u)$ to the mirror-symmetric $g(u)$ is
equivalent to considering the field on the part of the Minkowski plane not to
the right but to the left of trajectory $f(u)$ with the previous boundary
condition on the mirror.

The mean number of particles formed by a two-sided infinitely thin mirror on
the left part of the Minkowski plane is the same as on the right, since the integral
\begin{equation}
N=\int\!\!\int\limits_0^\infty
\frac{d\omega d\omega'}{(2\pi)^2}\,\vert\beta_{\omega'\omega}\vert^2  
\end{equation}
does not change when $\beta_{\omega'\omega}$ is replaced
by $\mp\beta_{\omega'\omega}$.  At the same time, the energy
\begin{equation}
{\cal E}'=\int\!\!\int\limits_0^\infty \frac{d\omega d\omega'}{(2\pi)^2}\,
\omega'\,\vert\beta_{\omega'\omega}\vert^2,                      
\end{equation}
emitted by the mirror to the left, generally speaking, is not equal to the
energy
\begin{equation}
{\cal E}=\int\!\!\int\limits_0^\infty \frac{d\omega
d\omega'}{(2\pi)^2}\, \omega\,\vert\beta_{\omega'\omega}\vert^2,    
\end{equation}
emitted to the right.

The equality of the mean numbers of particles emitted by a two-sided accelerated
mirror to the right and to the left suggests that the particles are generated in pairs
and fly off in opposite directions.  Quantity (5) is usually considered as the
mean number of actual quanta with frequency $\omega$ in the interval $d\omega$,
emitted to the right when a quantum with frequency $\omega'$ in the interval
$d\omega'$ is absorbed from the vacuum from the right.  The question arises
whether it is not possible to regard the same quantity as the mean number of pairs
of quanta emitted to the right and to the left with frequencies $\omega$ and
$\omega'$ in the intervals $d\omega$ and $d\omega'$, respectively.  In other
words, is $N^{-1}|\beta_{\omega'\omega}|^2$ the two-dimensional probability
distribution of frequencies $\omega$ and $\omega'$ of two quanta escaping to the
right and to the left with momenta $\omega$ and $-\omega'$?

Such an interpretation of the frequency distribution of bosons (fermions) emitted
by a mirror in 1 + 1 space would make the coincidence of this distribution with
the radiation spectrum of an electric (scalar) charge in 3 + 1 space detected in
Refs.~[1] and~[2] less formal.  Although, in the case of
mirror emission, the random quantities are the frequencies $\omega$ and
$\omega'$ of two bosons (fermions) escaping in different directions, while
in the case of charge emission, the random quantities are the components $k_+$
and $k_-$ of the wave vector of one vector (scalar) quantum emitted to the
right or to the left, corresponding to the sign of $k_+-k_->0$ or $<0$.

Let us give two more evidences of left--right symmetry of the wave field of an
accelerated mirror that are reflected by the Bogolyubov coefficients.

First, Eqs.~(6) and (7) or (9) and (10), obtained for the field to the right of the
mirror for the Bogolyubov coefficients, represent $|\beta_{\omega'\omega}|^2$
by a double integral over the entire $uv$ plane, as shown by Ref.~[2]. Thus
\begin{equation}
\vert\beta^B_{\omega'\omega}\vert^2=-\hbox{Re}\int\!\!\!\int\limits_{-\infty}
^\infty dudv\,\hbox{exp}\,[i\omega u+i\omega' f(u)-i\omega' v-i\omega g(v)],
\end{equation}
while $|\beta_{\omega'\omega}^F|^2$ differs from Eq.~(16)
by an additional factor of $-\sqrt{f'(u)g'(v)}$ under the integral.  Similarly,
in the double integral for the mean number of particles emitted to the right,
$$
N^{B,F}=\frac{1}{4\pi^2}\int\limits_{-\infty}^\infty du\,K^{B,F}(u),
$$
\begin{equation}
K^B(u)=\hbox{P}\int\limits_{-\infty}^\infty
\frac{dv}{v-f(u)}\left[\frac 1{g(v)-u}- \frac{f'(u)}{v-f(u)}\right], 
\end{equation}
\begin{equation}
K^F(u)=-\sqrt{f'(u)}\int\limits_{-\infty}^\infty\frac{dv}{v-f(u)}\left[
\frac{\sqrt{g'(v)}}{g(v)-u}-\frac{\sqrt{f'(u)}}{v-f(u)}\right],   
\end{equation}
the integration is carried out over the entire $uv$ plane, i.e.,
over all of Minkowski space, and not over the part of it lying to the right of
the trajectory.  The wave fields to the right and to the left of the trajectory
that satisfy the same condition on it are described by a single analytical
function and therefore are not independent.  Therefore, the frequencies of the
quanta emitted to the right and to the left are also not independent.

Second, the mean energies ${\cal E}$ and ${\cal E}'$ emitted to the right and to
the left, according to Ref.~[2], can be represented as integrals over the
proper time $\tau$ of the mirror:
\begin{equation}
{\cal E}^B=\frac 1{12\pi}\int\limits_{-\infty}^\infty [d\tau\,a^2\sqrt{f'}-
d(a\sqrt{f'})],                                         
\end{equation}
\begin{equation}
{\cal E}'^{B}=\frac 1{12\pi}\int\limits_{-\infty}^\infty\left[d\tau\frac{a^2}
{\sqrt{f'}}+d\left(\frac a{\sqrt{f'}}\right)\right],    
\end{equation}
Here $a$ is the acceleration of the mirror in its proper system.

The first terms under the integral in Eqs.~(19) and (20) represent the energy
irreversibly emitted by the mirror respectively to the right and to the left of the
section $d\tau$ of the trajectory.  In the mirror's proper system, these portions of
the energy are identical and equal $a^2d\tau/12\pi$, whereas the portions of
irreversibly emitted momentum equal $\pm a^2d\tau/12\pi$.  In the laboratory
system, these portions of the energy, because of the opposite directions of their
motion with respect to the velocity $\beta$ of the source, acquire Doppler factors
$\sqrt{f'}$ and $1/\sqrt{f'}$.  We recall that $\sqrt{f'}=\sqrt{(1+\beta)/(1-
\beta)}$.  The second, Schott terms under the integrals in Eqs.~(19) and (20)
``smear'' the formation region of radiation, as a result of which, for the
formation of radiation of energy, such intervals $\Delta\tau$ on which the
irreversibly emitted energy exceeds the change of the Schott energy are
substantial; i.e.,
\begin{equation}
\Delta\tau\,a^2\sqrt{f'}>\vert a\sqrt{f'}\vert,\qquad
\Delta\tau\frac{a^2}{\sqrt{f'}}>\vert\frac a{\sqrt{f'}}\vert,     
\end{equation}
or $\Delta\tau>a^{-1}$; the proper time interval must be greater than the inverse
proper acceleration.  The proper acceleration determines the characteristic
frequency of the radiation in the proper system and its scatter:
$\omega\sim\Delta\omega\sim a$.  Therefore, the condition $\Delta\tau a>1$ is
equivalent to the indeterminacy relation $\Delta\tau\Delta\omega>1$.

\section{Symmetry and the relations of certain integral quantities}
The following representations for the mean number $N$ of emitted particles and
the mean emitted energies ${\cal E}={\cal E}_+$ and ${\cal E}'={\cal E}_-$ are
convenient for explaining their properties with respect to certain space--time
transformations:
\begin{equation}
N^B=\int\!\!\int\limits_{-\infty}^\infty dudv\,S(u,v),\qquad
N^F=-\int\!\!\int\limits_{-\infty}^\infty dudv\,\sqrt{f'(u)g'(v)}S(u,v), 
\end{equation}
\begin{equation}
{\cal E}^B_{\pm}=\int\!\!\int\limits_{-\infty}^\infty dudv\,A_{\pm}(u,v),\qquad
{\cal E}^F_{\pm}=-\int\!\!\int\limits_{-\infty}^\infty dudv\,\sqrt{f'(u)g'(v)}
A_{\pm}(u,v),                                                            
\end{equation}
where $S$ and $A_\pm$ are singular functions $(\varepsilon,\delta\to+0)$:
\begin{equation}
S(u,v)=\frac 1{8\pi^2}\left[\frac{1}{(v-f(u)-i\varepsilon )(g(v)-u-i\delta)}+
\hbox{c.c.}\right],                                                       
\end{equation}
\begin{equation}
A_+(u,v)=\frac 1{8\pi^2i}
\left[\frac{1}{(v-f(u)-i\varepsilon)(g(v)-u-i\delta)^2}-\hbox{c.c.}\right],
\end{equation}                                                           
\begin{equation}
A_-(u,v)=\frac{1}{8\pi^2i}\left[\frac{1}{(v-f(u)-i\varepsilon)^2(g(v)-u-
i\delta)}-\hbox{c.c.}\right].                                            
\end{equation}

1.~Lorentz transformations.  The quantities $S(u,v)$, $\sqrt{f'(u)g'(u)}$, and
$dudv$ are scalars with respect to the Lorentz transformations, while
$A_\pm(u,v)$ transform as the $\pm$ components of a vector.  Therefore,
$N^{B,F}$ are Lorentz invariants, while ${\cal E}_\pm^{B,F}$ are the $\pm$
components of a vector.

2.~Mirror symmetry.  When the trajectory is replaced by a mirror-symmetric
trajectory, $f(u)\to g(u)$, $g(v)\to f(v)$, the integrals $N[f]$ and ${\cal
E}_\pm[f]$ transform, respectively, to
\begin{equation}
N[g]=N[f],\qquad  {\cal E}_\pm [g]={\cal E}_\mp [f],               
\end{equation}
since, for such a replacement,
\[
S(u,v)\to S(v,u),\quad A_\pm (u,v)\to A_\mp (v,u),\quad \sqrt{f'(u)g'(v)}\to
\sqrt{g'(u)f'(v)},
\]
after which the transformed integrals $N$ and ${\cal E}_\pm$ differ from the
untransformed $N$ and ${\cal E}_\mp$ only in the designation of the variables
of integration.  Thus, the mean numbers of particles emitted from the same
trajectory to the right and to the left are identical and do not change when the
trajectory is replaced by the mirror-symmetric one, while the mean energies
emitted to the right and to the left are different and transform into each other
when such replacement is made.

3.~Synchromirror transformation.  This discrete transformation consists of
replacing coordinates $u$ and $v$ with the coordinates
\begin{equation}
\tilde u=g(v),\;\tilde v=f(u),\;\;{\rm so\;that}\quad
u=g(\tilde v),\;\;v=f(\tilde u),
\end{equation}                                                        
Points $(u,v)$ and $(\tilde u,\tilde v)$, which are related
by transformation~(28), lie on the Minkowski plane on different sides of the
trajectory of the mirror on the intersection of the light cones whose vertices
are found on the trajectory at points
$A\mbox{\boldmath$($}u,f(u)\mbox{\boldmath$)$}$ and
$B\mbox{\boldmath$($}g(v),v\mbox{\boldmath$)$}$.  The points of any
compact region lying to the right of the trajectory are mapped one-to-one into the
points of a compact region lying to the left of the trajectory.

Functions $S(u,v)$ and $A_\pm(u,v)$ are form-invariant relative to
transformation~(28); i.e., their functional dependences on the new and old
variables are identical:
\begin{equation}
S(u,v)\equiv S\left(g(\tilde v),f(\tilde u)\right)=S(\tilde u,\tilde v),\qquad
A_\pm(u,v)=A_\pm(\tilde u,\tilde v).                                    
\end{equation}
Since the area element $dudv\sqrt{f'(u)g'(v)}$ appearing in the Fermi integrals
$N^F$ and ${\cal E}_\pm^F$ is also form-invariant relative to
transformation~(28); i.e.,
\begin{equation}
dudv\sqrt{f'(u)g'(v)}=d\tilde u d\tilde v\sqrt{f'(\tilde u)g'(\tilde v)},
\end{equation}
the contributions to the Fermi intervals from any two regions on the $uv$ plane
related by symmetry transformation (28) are identical.  In particular, the
contributions from the entire region to the right and the entire region to the left
of the trajectory are identical.

In the Bose integrals $N^B$ and ${\cal E}_\pm^B$, the contributions of the
 right-hand and left-hand regions related by transformation~(28) are, generally
speaking, different, since the area element $dudv$ that appears in these integrals,
unlike the functions $S$ and $A_\pm$ being integrated, is mapped by
transformation~(28) into the unequal element $d\tilde ud\tilde v$:
\begin{equation}
dudv=d\tilde ud\tilde vf'(\tilde u)g'(\tilde v),\qquad                
d\tilde ud\tilde v=dudvf'(u)g'(v).
\end{equation}
Therefore, the contributions to the Bose integrals from these two elementary
areas are proportional to their areas; i.e., their ratio equals the Jacobian of the
transformation.

Transformation~(28) of the variables of integration of course does not change the
values of the integrals $N$ and ${\cal E}_\pm$.  Its meaning is that the local
contributions to $N$ and ${\cal E}_\pm$ from any pair of
 right-hand and left-hand regions associated by transformation~(28) have a
definite symmetry or asymmetry.  Namely, for Fermi integrals, this symmetry
consists of the equality of such contributions, whereas, for Bose integrals, it
consists of left--right asymmetry of the contributions, determined by the Jacobian
of the transformation.

\section{Radiation of two-sided mirror, quantum approach}
For a consistent description of the quantum wave field lying both to the right and
to the left of the mirror and satisfying a single condition on the mirror, it is
convenient to use the two complete sets $\{\phi_{out\,\omega},\phi_{out\,\omega}
^*\}$ and $\{\phi_{in\,\omega'},\phi_{in\,\omega'}^*\}$ of solutions of the wave
equation, given in Refs.~[1] and [2]. Possessing in the right-hand Minkowski
plane the physical meaning of the {\it out\/} and {\it in\/} sets and satisfying
the boundary condition on the mirror, these solutions can be smoothly extended
into the left half-plane with no change of their functional form.  However, in
the left half-plane, these sets acquire the physical meaning of the {\it in\/}
and {\it out\/} sets, respectively, and they must be designated as
$\{\phi_{in\,\omega},\phi_{in\,\omega}^*\}$ and
$\{\phi_{out\,\omega'},\phi_{out\,\omega'}^*\}$.

Each such solution is actually unambiguously characterized by the frequency
$\omega$ or $\omega'$ of its monochromatic component travelling to the right
or to the left and by the condition on the mirror.  For a Lorentzian
transformation with velocity $\beta$ along the $x$ axis, the frequencies $\omega$
and $\omega'$ transform into $\tilde\omega$ and $\tilde\omega'$ according to the
mutually inverse laws
\begin{equation}
\tilde\omega=D^{-1}(\beta)\,\omega,\quad \tilde\omega'=D(\beta)\,\omega',\quad
D(\beta)=\sqrt{\frac{1+\beta}{1-\beta}},                               
\end{equation}
where $D(\beta)$ is the Doppler factor.  $\omega$ and $\omega'$ thus possess
the opposite covariance.  Below, frequencies that transform like $\omega$ will be
equipped with an even number of primes, while those that transform like
$\omega'$ will have an odd number.  Then the subscript {\it in\/} or {\it out}, in
addition to the frequency, will simply indicate the side of the Minkowski plane on
which the solution is considered.

The expansion of the solutions of the first set in the solutions of the second set
and the inverse expansion have been written by us (in the right-hand half-plane)
in the form
\begin{equation}
\phi_{out\,\omega}=\alpha_{\omega'\omega}\phi_{in\,\omega'}+
\beta_{\omega'\omega}\phi_{in\,\omega'}^*,                           
\end{equation}
\begin{equation}
\phi_{in\,\omega'}=\alpha_{\omega'\omega}^*\phi_{out\,\omega}\mp
\beta_{\omega'\omega}\phi_{out\,\omega}^*,                            
\end{equation}
or, if matrix notation is used,
\begin{equation}
\left(\begin{array}{c}
\phi_{out}\\
\phi_{out}^*
\end{array}\right)=
\left(\begin{array}{cc}
\tilde\alpha&\tilde\beta\\
\beta^+&\alpha^+
\end{array}\right)
\left(\begin{array}{c}
\phi_{in}\\
\phi_{in}^*
\end{array}\right),\quad
\left(\begin{array}{c}
\phi_{in}\\
\phi_{in}^*
\end{array}\right)=
\left(\begin{array}{cc}
\alpha^*&\mp\beta\\
\mp\beta^*&\alpha
\end{array}\right)
\left(\begin{array}{c}
\phi_{out}\\
\phi_{out}^*
\end{array}\right).                                            
\end{equation}
As a consequence of the orthogonality and normalization of the solutions in
both sets, the matrices that appear in Eqs.~(35) are mutually inverse.  This
means that the Bogolyubov coefficients satisfy four independent matrix
relations:
\begin{equation}
\begin{array}{cc}
\alpha^+\alpha\mp\beta^+\beta=1,& \beta^+\alpha^*\mp\alpha^+\beta^*=0,\\
\alpha\alpha^+\mp\beta^*\tilde\beta=1,& \alpha\beta^+\mp\beta^*\tilde\alpha=0.
\end{array}                                                     
\end{equation}
On the left-hand half-plane, Eqs.~(33)--(35) are conserved, but a new physical
meaning requires the interchange of the subscripts $in\rightleftharpoons out$
in the functions, which is equivalent to transformation~(12).

For a quantized field in the right half-plane, the connection of the {\it in\/}
and {\it out\/} absorption and creation operators $a$ and $a^+$ is given by the
Bogolyubov transformations
\begin{equation}
\left(\begin{array}{c}
a_{in}\\
a_{in}^+
\end{array}\right)=
\left(\begin{array}{cc}
\alpha&\beta^*\\
\beta&\alpha^*
\end{array}\right)
\left(\begin{array}{c}
a_{out}\\
a_{out}^+
\end{array}\right),\quad
\left(\begin{array}{c}
a_{out}\\
a_{out}^+
\end{array}\right)=
\left(\begin{array}{cc}
\alpha^+&\mp\beta^+\\
\mp\tilde\beta&\tilde\alpha
\end{array}\right)
\left(\begin{array}{c}
a_{in}\\
a_{in}^+
\end{array}\right).                                           
\end{equation}
For a field in the left-hand half-plane, an interchange of the subscripts
$in\rightleftharpoons out$ is required on operators $a$ and $a^+$ in
transformations (37).  This again is equivalent to transformation~(12).

Following DeWitt's paper [3] and its notation, we represent the vector of the
vacuum state of the field in the distant past in the form of an expansion in the
vectors of the $n$-particle states of the field in the distant future:
\begin{equation}
\vert in\rangle=e^{iW}\sum_{n=0}^\infty\frac{i^{n/2}}{n!}\sum_{i_1i_2\ldots
i_n} V_{i_1i_2\ldots i_n}\vert i_1i_2\ldots i_n out\rangle.        
\end{equation}
In our case, by the quantum numbers $i_1i_2\ldots i_n$ of
the {\it out\/} states of the individual particles should be understood
frequencies that transform like $\omega$ or like $\omega'$ if one is dealing
with the field, respectively, to the right or to the left of the mirror.

Using the equation $a_{in}|in\rangle=0$, transformations (37), and the expansion
given by Eq.~(38), it is easy to show [3,4] that the relative production
amplitudes $V_{i_1i_2\ldots i_n}$ of $n$ particles equal zero for odd $n$,
whereas, for even $n$, they are expressed in terms of the production amplitude
of a pair of particles:
\begin{equation}
V_{i_1i_2\ldots i_n}=\sum_p\delta_p V_{i_1i_2}V_{i_3i_4}\ldots V_{i_{n-1}i_n}.
\end{equation}
Here $\sum\limits_p$ denotes summation over $n!/2^{n/2}(n/2)!$ different
pairings of subscripts $i_1i_2\ldots i_n$, while $\delta_p=1$ for bosons and
$\delta_p=\pm1$ for fermions when the permutation leading to the given pairing
is, respectively, even or odd.  The production amplitudes of a pair of particles
with frequencies $\omega''$ and $\omega$ in the right-hand region and
frequencies $\omega'''$ and $\omega'$ in the left-hand region equal
\begin{equation}
V_{\omega''\omega}=i(\alpha^{-1}\beta^*)_{\omega''\omega},\qquad
V_{\omega'''\omega'}=-i(\beta\alpha^{-1})^*_{\omega'''\omega'}.       
\end{equation}
They are related to each other by transformation~(12), which is symmetric for a
Bose field and antisymmetric for a Fermi field, as follows from Eqs.~(36).

The indicated number of terms in the amplitude of Eq.~(39) appears in
connection with its symmetrization (antisymmetrization) and equals the number
$n!$ of permutations of its subscripts, reduced by a factor of $2^{n/2}$ because
of the already existing symmetry (antisymmetry) of the two-particle amplitudes
and by a factor of $(n/2)!$ because of the inessentiality of permutations of
these amplitudes.

Particle production in pairs is explained by the linearity of the Bogolyubov
transformations in the operators $a$ and $a^+$.  Operator $a_{in}$, when it acts
on the $n$-particle {\it out\/} state, transforms it into a superposition
of $n-1$-particle and $n+1$-particle {\it out\/} states.  Therefore, in the
expansion of the null vector $a_{in}|in\rangle$ in the $n$-particle {\it out\/}
states, the expansion coefficients equal to zero represent the linear relation
between the amplitudes of the $n+1$-particle and $n-1$-particle creations.
Since $n\geq0$, the amplitude of the single-particle production $V_{i_1}$ is
equal to zero, and, along with it, all the formation amplitudes of an odd number
of particles.

The absolute amplitudes of the $n$-particle production are determined and are
related to the relative amplitudes by
\begin{equation}
\langle out\,i_1i_2\ldots i_n\vert in\rangle\equiv\langle out\vert a_{out\,i_n}
\ldots a_{out\,i_2}a_{out\,i_1}\vert in\rangle=e^{iW}i^{n/2} V_{i_1i_2\ldots
i_n}.                                                     
\end{equation}

The vacuum-conservation amplitude $\langle out|in\rangle=e^{iW}$ is
determined to within a phase factor by the fact that the total probability of the
transition from the initial vacuum state is equal to one:
\begin{equation}
1=\sum_{n=0}^\infty\frac{1}{n!}\sum_{i_1i_2\ldots i_n}\vert\langle out\,i_1i_2
\ldots i_n\vert
in\rangle\vert^2=e^{-2\,{\rm Im}\,W}\sum_{n=0}^\infty\frac1{n!}
\sum_{i_1i_2\ldots i_n}\vert V_{i_1i_2\ldots i_n}\vert^2.               
\end{equation}
The sum of the relative probabilities
\begin{equation}
q_n=\frac1{n!}\sum_{i_1i_2\ldots i_n}\vert V_{i_1i_2\ldots i_n}\vert^2    
\end{equation}
of the production of $n$ particles (or of $n/2$ pairs) on the right-hand side of
Eq.~(42) we shall call the statsum.  It can be shown that, in the case
considered here, in which pairs of identical particles and antiparticles are
formed, the statsum equals
\begin{equation}
\sum_{n=0}^\infty\frac1{n!}\sum_{i_1i_2\ldots i_n}\vert V_{i_1i_2\ldots i_n}
\vert^2=\det\,(1\mp M)^{\mp1/2}=\exp\,\left(\mp\frac12\hbox{tr}\,
\ln(1\mp M)\right),                                                     
\end{equation}
where $M=VV^+$ is a Hermitian positive-semidefinite matrix formed from the matrices in Eqs.~(40).
In particular, the first three terms of the statsum, determined by the relative
amplitudes
\begin{equation}
1,\quad V_{i_1i_2},\quad V_{i_1i_2}V_{i_3i_4}\pm V_{i_1i_3}V_{i_2i_4}+
V_{i_1i_4}V_{i_2i_3},                                                   
\end{equation}
and by Eq.~(43), are equal, respectively, to
\begin{equation}
q_0=1,\quad q_2=\frac12\hbox{tr}\,M,\quad q_4=\frac18 (\hbox{tr}\,M)^2\pm
\frac14\hbox{tr}\,M^2.                                                  
\end{equation}
The absolute probabilities of forming $n$ pairs are equal to
$p_{2n}=p_0q_{2n}$, where $p_0$ is the vacuum-conservation probability:
\begin{equation}
p_0=e^{-2\,{\rm Im}\,W},\qquad 2\,{\rm Im}\,W=\mp\frac12\hbox{tr}\,\ln(1\mp M).
\end{equation}                                                         
Since the relative probabilities $q_{2n}(M)$ of producing $n$ pairs are
homogeneous functions of degree $n$, $q_{2n}(\lambda M)=\lambda^nq_{2n}(M)$,
it is convenient to compute the mean number of pairs from
\begin{equation}
\bar n=\sum_{n=0}^\infty np_{2n}=p_0\lambda\frac{\partial}{\partial\lambda}
\sum_{n=0}^\infty\lambda^nq_{2n}(M)\vert_{\lambda=1}=\lambda\frac{\partial}
{\partial\lambda}2\,\hbox{Im}\,W(\lambda M)\vert_{\lambda=1}=\frac12\hbox{tr}\,
\frac{M}{1\mp M}.                                                      
\end{equation}
Matrices $M$ are different for the right-hand and left-hand regions:
\begin{equation}
M=VV^+=\beta^+\beta(1\pm\beta^+\beta)^{-1},                            
\end{equation}
\begin{equation}
M=VV^+=\beta^*\tilde\beta(1\pm\beta^*\tilde\beta)^{-1},                 
\end{equation}
but are related to each other by transformation~(12).  However, the
positive-definite quantities ${\rm tr}M^n$, $n=1,2,\ldots$, are invariants of
this transformation.  Therefore, the total probabilities given above for
conservation of the vacuum, $p_0$, and of the production of $n$ pairs,
$p_{2n}$, and the mean number of pairs, $\overline n$, are identical for the
right-hand and left-hand regions.  In particular, the quantities
\begin{equation}
p_0=e^{-2\,{\rm Im}\,W},\qquad 2\,{\rm Im}\,W=\pm\frac12\hbox{tr}\,\ln(1\pm
\beta^+\beta),                                                        
\end{equation}
\begin{equation}
p_2=e^{-2\,{\rm Im}\,W}\frac12\hbox{tr}\,\beta^+\beta(1\pm\beta^+\beta)^{-1},
\end{equation}                                                        
\begin{equation}
\bar n=\frac12\hbox{tr}\,\beta^+\beta                                 
\end{equation}
do not change under transformation (12) or
$\beta^+\beta\rightarrow\beta\beta^+$.

Nevertheless, the frequency distributions of the probabilities and of the mean
number of particles possess no left--right symmetry.  Thus, the production
probability of one pair, one particle of which has a definite frequency while the
other has any frequency, equals
\begin{equation}
p_{2\omega}=e^{-2\,{\rm Im}\,W}\left(\frac{\beta^+\beta}{1\pm\beta^+\beta}\right)
_{\omega\omega},                                                       
\end{equation}
for the right-hand region and equals
\begin{equation}
p'_{2\omega'}=e^{-2\,{\rm Im}\,W}\left(\frac{\beta\beta^+}{1\pm\beta\beta^+}
\right)_{\omega'\omega'}.                                               
\end{equation}
for the left-hand region. The frequency distributions of the mean number of
particles emitted by the mirror to the right and to the left are also functionally
different from each other:
\begin{equation}
N_\omega=(\beta^+\beta)_{\omega\omega},\qquad
N'_{\omega'}=(\beta\beta^+)_{\omega' \omega'}.                          
\end{equation}

Along with the amplitudes given by Eq.~(41) for particle production from
vacuum by the mirror, it is necessary to consider the amplitudes of single-particle
scattering by the mirror
\begin{equation}
\langle out\,\omega\vert\omega'\,in\rangle=\langle out\vert a_{out\,\omega}
a_{in\,\omega'}^+\vert in\rangle=e^{iW}\alpha_{\omega\omega'}^{-1},       
\end{equation}
\begin{equation}
\langle out\,\omega'\vert\omega\,in\rangle=\langle out\vert a_{out\,\omega'}
a_{in\,\omega}^+\vert in\rangle=e^{iW}\alpha_{\omega\omega'}^{-1*},       
\end{equation}
for the right-hand and left-hand regions, respectively.  These amplitudes differ
only in their phases.  Of course, they are related to each other by
transformation~(12), but we shall be interested in their relation to the
corresponding pair-production amplitudes:
\begin{equation}
\langle out\,\omega''\omega\vert in\rangle=
-e^{iW}(\alpha^{-1}\beta^*)_{\omega''\omega}=-\sum_{\omega'}\langle
out\,\omega''\vert\omega'\,in\rangle\beta_{\omega'\omega}^*,          
\end{equation}
\begin{equation}
\langle out\,\omega'\omega'''\vert in\rangle=
e^{iW}(\beta\alpha^{-1})_{\omega'\omega'''}^*=\sum_\omega\beta_{\omega'\omega}^*
\langle out\,\omega'''\vert\omega\,in\rangle.                          
\end{equation}

Since the pair-production amplitudes and the single-particle scattering amplitudes
are quantities that can in principle be experimentally measured from the
corresponding probabilities, Eqs.~(59) and (60) make it possible to
experimentally measure $\beta_{\omega'\omega}^*$.  Moreover, these
relationships make it possible to regard $\beta_{\omega'\omega}^*$ as the
amplitude of the source of a pair of particles potentially emitted to the right and
to the left with frequencies $\omega$ and $\omega'$, respectively.  In this case, if
a particle with frequency $\omega$ actually escaped to the right, a particle with
frequency $\omega'$ does not escape to the left, but experiences an internal
reflection and is actually emitted to the right with altered frequency $\omega''$.
Conversely, if a particle with frequency $\omega'$ actually escaped to the left, a
particle with frequency $\omega$ cannot escape to the right, but, after internal
reflection, is actually emitted to the left with another frequency $\omega'''$.

For fermions, amplitude $\beta_{\omega'\omega}^F$ is diagonal in the
projection of the spin of the {\it in\/} and {\it out\/} waves (see
Ref.~[2]).  But one of the waves forming $\beta_{\omega'\omega}^F$
has a negative frequency and therefore describes an antiparticle with frequency
and spin projection opposite in sign to the frequency and spin projection of this
wave (see $\S26$ in Ref.~[5] or $\S9$ of chap.~2 in Ref.~[6]).
Thus, the spin of a pair of generated fermions equals zero.
This is confirmed by the scalar nature of the identically equal integrals in
Eqs.~(9) and (10), in which $du\sqrt{f'(u)}$ and $dv\sqrt{g'(v)}$ are elements
of proper time $d\tau$, and by their coincidence,
\begin{equation}
\beta_{\omega'\omega}^{F*}=\frac1e\rho\,(k_+,k_-)                   
\end{equation}
with the Fourier component of the scalar-charge density in 3 + 1 space.

Amplitude $\beta_{\omega'\omega}^{B*}$ of the source of a boson pair,
according to Eqs.~(6) and (7), is linearly expressed in terms of the Fourier
components $j_\pm(k)$ of the current density of an electric charge in 3 + 1
space:
\begin{equation}
\beta_{\omega'\omega}^{B*}=-\sqrt{\frac{k_+}{k_-}}\frac{j_-}{e}=        
\sqrt{\frac{k_-}{k_+}}\frac{j_+}{e},
\end{equation}
\begin{equation}
j_-=e\int_{-\infty}^\infty du\,\exp\left[\frac i 2 (k_+u+k_-f(u))\right],
\quad
j_+=e\int_{-\infty}^\infty dv\,\exp\left[\frac i 2 (k_-v+k_+g(v))\right],
\end{equation}
see also Eqs.~(1) and (2) in this paper and Eqs.~(43) and (44) in Ref.~[1].
The last equality in Eq.~(62) is none other than the current-transverseness
condition, $k_+j_-+k_-j_+=0$.  It can also be seen from Eq.~(62) that
$\beta_{\omega'\omega}^B$ is a pseudoscalar, since, at the reflection
$k_\pm\to k_\mp$, $j_\pm\to j_\mp$, and $\beta^B$ changes sign.  Vector
$j_\alpha(k)$ is spacelike and, in a system where $k_+=k_-$ (or
$\omega=\omega'$), has only a spatial component, precisely equal to
$e\beta_{\omega'\omega}^B$.  In covariant form,
\[
e\beta_{\omega'\omega}^{B*}=\varepsilon_{\alpha\beta}k^\alpha j^\beta /
\sqrt{k_+k_-}.
\]
Thus, the source of a boson pair is the conserved current vector given by
Eqs.~(63), and this means that the spin of a pair equals 1, see [7].

The fact that the spin of a boson pair equals 1 while that of a fermion pair
equals 0 is essential for understanding the coincidence of the spectra of a
mirror and of a charge.

If $\beta_{\omega'\omega}^*$ is small, i.e., if the mean number of emitted
quanta is small, then, as is easy to obtain from Eqs.~(6) and (9),
\begin{equation}
\alpha_{\omega'\omega}\approx 2\pi\delta(\tilde\omega'-\tilde\omega),\qquad
\alpha_{\omega\omega'}^{-1}\approx 2\pi\delta(\tilde\omega-\tilde\omega'),
\end{equation}
where $\tilde\omega$ and $\tilde\omega'$ are related to $\omega$ and $\omega'$
by transformation (32), in which $\beta$ is the effective velocity of the mirror on
the emission section.  In this approximation, the emission amplitudes given by
Eqs.~(59) and (60) for pairs of particles with frequencies $\omega$ and
$\omega''$ to the right and pairs of particles with frequencies of $\omega'$ and
$\omega'''$ to the left equal, respectively,
\begin{equation}
\langle out\,\omega''\omega\vert in\rangle\approx
-e^{iW}D^{-1}(\beta)\,\beta_{\omega'\omega}^*,\qquad
\omega'=D^{-2}(\beta)\,\omega'',                                     
\end{equation}
\begin{equation}
\langle out\,\omega'\omega'''\vert in\rangle\approx
e^{iW}D(\beta)\,\beta_{\omega'\omega}^*,\qquad
\omega=D^2(\beta)\,\omega'''.                                        
\end{equation}
These formulas, including the connection between the frequencies of the waves
incident on the mirror and reflected from it, confirm the interpretation of
$\beta_{\omega'\omega}^*$ given above.

We now turn our attention to interference effects in the production of Bose and
Fermi particles.  They become most substantial when matrices $M$ for bosons
and fermions satisfy the conditions
\[
\mp\frac12\hbox{tr}\,\ln(1\mp M)=\mp\ln\left(1\mp\frac12\hbox{tr}\,M\right),
\]
i.e.,
\begin{equation}
\frac12\hbox{tr}\,M^n=\left(\frac12\hbox{tr}\,M\right)^n,\quad n=2,3,\ldots .
\end{equation}                                                        
Then the statsum given by Eq.~(44) for Bose and Fermi particles reduces,
respectively, to
\begin{equation}
\left(1-\frac12\hbox{tr}\,M\right)^{-1}\qquad {\rm and}\qquad
1+\frac12\hbox{tr}\,M.                                                
\end{equation}
This means that the probabilities of producing $n$ pairs of bosons form the
geometrical progression
\begin{equation}
p_{2n}^B=p_0^Bq_2^{Bn},\quad p_0^B=1-\frac12\hbox{tr}\,M,\quad
q_2^B=\frac12\hbox{tr}\,M,
\end{equation}
while the probabilities of emitting two or more pairs of fermions disappear; i.e.,
only the production of one fermion pair is possible:
\begin{equation}
p_0^F=\left(1+\frac12\hbox{tr}\,M\right)^{-1},\quad p_2^F=p_0^F\frac12
\hbox{tr}\,M,\quad p_{2n}^F=0,\quad n\ge 2.                             
\end{equation}
In other words, the conditions given by Eqs.~(67) denote the most constructive
interference of bosons and the most destructive interference of fermions.  In
these cases, the mean-square fluctuation of number of boson pairs is always
greater than $\bar n^B $, while that of the fermion pairs is less than
$\bar n^F $, being equal to $\bar n(1\pm\bar n)$, where
\begin{equation}
0<\bar n^B=\frac{\frac12\hbox{tr}\,M}{1-\frac12\hbox{tr}\,M}<\infty,\qquad
0<\bar n^F=\frac{\frac12\hbox{tr}\,M}{1+\frac12\hbox{tr}\,M}<1.  
\end{equation}
We are less interested in the case in which interference effects can be
neglected.  In this case,
\begin{equation}
\hbox{tr}\,M^k\ll\hbox{tr}\,M,\;1\:;\qquad k\ge 2,                    
\end{equation}
and the probability distribution over the number of generated pairs coincides
with the Poisson distribution:
\begin{equation}
p_{2n}=e^{-\bar n}\frac{(\bar n)^n}{n!},\qquad
\bar n=\frac12\hbox{tr}\,\beta^+\beta.                                  
\end{equation}

\section{Emission of pairs consisting of nonidentical particles and
antiparticles}

In the case of pair production of nonidentical particles and antiparticles
($ab$ pairs), the direct and inverse Bogolyubov transformations (37) are
replaced by
\begin{equation}
\left(\begin{array}{c}
a_{in}\\
b_{in}^+
\end{array}\right)=
\left(\begin{array}{cc}
\alpha_{aa}&\beta_{ab}^*\\
\beta_{ba}&\alpha_{bb}^*
\end{array}\right)
\left(\begin{array}{c}
a_{out}\\
b_{out}^+
\end{array}\right),\quad
\left(\begin{array}{c}
a_{out}\\
b_{out}^+
\end{array}\right)=
\left(\begin{array}{cc}
\alpha_{aa}^+&\mp\beta_{ba}^+\\
\mp\tilde\beta_{ab}&\tilde\alpha_{bb}
\end{array}\right)
\left(\begin{array}{c}
a_{in}\\
b_{in}^+
\end{array}\right).                                           
\end{equation}
These transformations contain not two but four matrices $\alpha_{aa},
\alpha_{bb}, \beta_{ab}$, and $\beta_{ba}$, which satisfy not the four Eqs.(36)
but the six equations
\begin{equation}
\begin{array}{cc}
\alpha_{aa}^+\alpha_{aa}\mp\beta_{ba}^+\beta_{ba}=1,&
\beta_{ba}^+\alpha_{bb}^*\mp\alpha_{aa}^+\beta_{ab}^*=0,\\
\alpha_{bb}^+\alpha_{bb}\mp\beta_{ab}^+\beta_{ab}=1,&
\alpha_{aa}\alpha_{aa}^+\mp\beta_{ab}^*\tilde\beta_{ab}=1,\\
\alpha_{bb}\alpha_{bb}^+\mp\beta_{ba}^*\tilde\beta_{ba}=1,&
\alpha_{aa}\beta_{ba}^+\mp\beta_{ab}^*\tilde\alpha_{bb}=0.
\end{array}                                                    
\end{equation}
However, these relationships can be written in the form of Eqs.(36) if
$\alpha$ and $\beta$ stand for the 2$\times$2 matrices consisting of the
indicated quarters:
\begin{equation}
\alpha =
\left(\begin{array}{cc}
\alpha_{aa}& 0\\
0& \alpha_{bb}
\end{array}\right),\quad
\beta =
\left(\begin{array}{cc}
0& \beta_{ab}\\
\beta_{ba}& 0
\end{array}\right).
\end{equation}                                                   

As can be seen from Eqs. (74), the interchange $in \rightleftharpoons out$ is
now equivalent to the interchange
\begin{equation}
\alpha_{aa}\to\alpha_{aa}^+,\quad \alpha_{bb}\to\alpha_{bb}^+,\quad
\beta_{ab}\to\mp\tilde\beta_{ba},\quad \beta_{ba}\to\mp\tilde\beta_{ab}, 
\end{equation}
which can be represented in the form of the transformation (12) if $\alpha$ and
$\beta$ stand for the matrices of Eqs.(76).

Using for the in-vacuum state an expansion of the type of Eq.(38) and the
equations $a_{in}\vert in \rangle = b_{in} \vert in \rangle =0$, it can be
shown that all the emission amplitudes of an odd number of particles equal
zero, while the production amplitudes of an even number of particles are
products of the production amplitudes of $ab$ pairs:
\begin{equation}
V_{\omega''\omega}^{ab}=i(\alpha_{aa}^{-1}\beta_{ab}^*)_{\omega''\omega},
\quad V_{\omega'''\omega'}^{ab}=-i(\beta_{ab}\alpha_{bb}^{-1})_{\omega'''
\omega'}^*,
\end{equation}
respectively for the right-hand and the left-hand regions.  As follows from
Eqs.~(75), the amplitudes given by Eqs.~(78) possess intrinsic Bose symmetry or
Fermi antisymmetry:
\begin{equation}
V_{\omega''\omega}^{ab}=\pm V_{\omega\omega''}^{ba}\equiv\pm i(\alpha_{bb}^{-1}
\beta_{ba}^*)_{\omega\omega''},\quad
V_{\omega'''\omega'}^{ab}=\pm V_{\omega'\omega'''}^{ba}\equiv\mp i(\beta_{ba}
\alpha_{aa}^{-1})_{\omega'\omega'''}^*.                                   
\end{equation}
Thus, the formation amplitude of an $ab$ pair can be denoted via $V_{i_1i_2}$,
where the subscript $i_1$ characterizes the state of the particle and $i_2$
that of the antiparticle. The production of two $ab$ pairs is described by the
amplitude
\begin{equation}
V_{i_1i_2i_3i_4}=V_{i_1i_2}V_{i_3i_4}\pm V_{i_3i_2}V_{i_1i_4},         
\end{equation}
symmetric (antisymmetric) separately with respect to states
$i_1$ and $i_3$ of the particles and separately with respect to states $i_2$
and $i_4$ of the antiparticles.  We also write the production amplitude of
three pairs:
$$
V_{i_1i_2i_3i_4i_5i_6}=V_{i_1i_2}V_{i_3i_4}V_{i_5i_6}\pm V_{i_3i_2}V_{i_1i_4}
V_{i_5i_6}+V_{i_3i_2}V_{i_5i_4}V_{i_1i_6}\pm $$
\begin{equation}
\pm V_{i_1i_2}V_{i_5i_4}V_{i_3i_6}+V_{i_5i_2}V_{i_1i_4}V_{i_3i_6}\pm V_{i_5i_2}
V_{i_3i_4}V_{i_1i_6}.
\end{equation}                                                         
In the general case, the production amplitude of $n/2$ pairs has the form
\begin{equation}
V_{i_1i_2\ldots i_n}=\sum_{p}\delta_{p}V_{i_1i_2}V_{i_3i_4}\ldots V_{i_{n-1}
i_n},                                                                   
\end{equation}
where the sum is taken over all $(n/2)!$ terms that differ
by a permutation of the odd subscripts (or, what is the same thing, by a
permutation of the even subscripts), with $\delta_p=\pm1$ in the case of
fermions for an even or odd permutation, respectively, while $\delta_p=1$ in
the case of bosons.  Then amplitude $V_{i_1i_2\ldots i_n}$ will be symmetric
(antisymmetric) both over particle states $i_1i_3\ldots i_{n-1}$ and over
antiparticle states $i_2i_4\ldots i_n$.

The relative probability
\begin{equation}
q_n=\frac{1}{(n/2)!(n/2)!}\sum_{i_1i_2\ldots i_n}\vert V_{i_1i_2\ldots i_n}
\vert^2                                                                  
\end{equation}
of producing $n/2$ pairs consisting of nonidentical particles and antiparticles
contains the factor $1/(n/2)!(n/2)!$, which, along with the symmetry
(antisymmetry) of amplitude $V_{i_1i_2\ldots i_n}$ separately for even and
separately for odd subscripts, makes it possible to sum over the particle and
antiparticle states, considering the ranges of variation of the quantum numbers of
these states to be independent.  Without this factor, the sum over $i_1i_2\ldots
i_n$ would have had to contain only physically different states.  In our case, for
example, it would be unambiguous that the frequencies of the particles must
satisfy the condition $\omega_1\geq\omega_3\geq\ldots\geq\omega_{n-1}$, while
the frequencies of the antiparticles must satisfy the condition
$\omega_2\geq\omega_4\geq\ldots\geq\omega_n$.

It is easy to construct the first four terms of the statsum in terms of the relative
amplitudes shown above:
$$
q_0=1,\quad q_2={\rm tr}\,M,\quad q_4=\frac12({\rm tr}\,M)^2\pm\frac12{\rm tr}
\,M^2,
$$
\begin{equation}
q_6=\frac16({\rm tr}\,M)^3\pm\frac12{\rm tr}\,M\,{\rm tr}\,M^2+
\frac13{\rm tr}\,M^3.                                                 
\end{equation}
For the statsum as a whole, we get
\begin{equation}
\sum_{n=0}^{\infty}\frac{1}{(n/2)!(n/2)!}\sum_{i_1i_2\ldots i_n}\vert V_{i_1i_2
\ldots i_n}\vert^2=\det (1\mp M)^{\mp 1}=\exp (\mp {\rm tr}\,\ln (1\mp M)).
\end{equation}                                                        
Here, as in Eq.~(44), $M=VV^+$ is a Hermitian positive-semidefinite matrix.
It is given by Eqs.~(49) and (50), in which by $\beta$ is meant, respectively,
$\beta_{ba}$ and $\beta_{ab}$.

Just as above, the absolute probabilities of the formation of $n$ pairs of
nonidentical particles and antiparticles equal $p_{2n}=p_0q_{2n}$, where $p_0$
is the vacuum-conservation probability:
\begin{equation}
p_0=e^{-2\,{\rm Im}\,W},\quad 2\,{\rm Im}\,W=\mp{\rm tr}\,\ln(1\mp M).  
\end{equation}
The mean number of pairs, computed according to the rule given in Eq.~(48),
equals
\begin{equation}
\bar n={\rm tr}\frac{M}{1\mp M}.                                        
\end{equation}
It can be seen that these formulas differ from the corresponding Eqs.~(47) and
(48) for pair production of identical particles by replacing (1/2)tr by tr in
the latter equations.  Because of the $a\rightleftharpoons b$ symmetry in
the matrices, under the tr sign, by $\beta$ can be understood both $\beta_{ba}$
and $\beta_{ab}$.

It is easy to see that this rule connects all the formulas for the integral
characteristics of pair production of identical particles with the formulas of the
corresponding characteristics of $ab$-pair production.  Thus, in order to obtain
from Eqs.~(51)--(53) and (67)--(73) the analogous expressions for $ab$-pair
production, it is sufficient to replace (1/2)tr in these formulas with tr and
by $\beta$ to understand $\beta_{ba}$ or $\beta_{ab}$.

As far as the spectral characteristics shown, for example, in Eqs.~(54)--(56)
are concerned, they undergo no changes when the transition is made to the case
under consideration, if by $\beta$ is understood $\beta_{ba}$ ($\beta_{ab}$)
for the spectrum of particles (antiparticles) emitted to the right and
$\beta_{ab}$  ($\beta_{ba}$) for the spectrum of particles (antiparticles)
emitted to the left.

In fact, for the differential probability $p_{2\omega}$ shown in Eq.~(54), the
original integral
\begin{equation}
p_{2\omega}=\int\limits_0^\infty\,\frac{d\omega''}{2\pi}\vert\langle out\,
\omega\omega''\vert in\rangle\vert^2                                   
\end{equation}
represents it as the sum of the probabilities of physically different events
regardless of whether the particles are identical or not.  However, the total
 pair-formation probability $p_2$ as a sum of probabilities of physically
different events for identical particles is represented by the integral
\begin{equation}
p_2=\int\limits_0^\infty\,\frac{d\omega}{2\pi}\int\limits_0^{\omega}\,\frac
{d\omega''}{2\pi}\vert\langle out\,\omega\omega''\vert in\rangle\vert^2=
\frac12\,\int\limits_0^\infty\,\frac{d\omega}{2\pi}p_{2\omega}.          
\end{equation}
since the states in this case differ only by the values of the large frequency
$\omega$ and the small frequency $\omega''$ of two identical particles.  At the
same time, for an $ab$ pair,
\begin{equation}
p_2=\int\limits_0^\infty\,\frac{d\omega}{2\pi}\int\limits_0^\infty\,\frac
{d\omega''}{2\pi}\vert\langle out\,\omega\omega''\vert in\rangle\vert^2=
\int\limits_0^\infty\,\frac{d\omega}{2\pi}p_{2\omega}.                   
\end{equation}
since the states differ in the frequencies $\omega''$ and $\omega$ of the particle
and antiparticle independently of each other, while the particle and antiparticle
differ in turn in that they interact differently with the counters.

Turning to the amplitude of $ab$-pair production,
\begin{equation}
\langle out\vert b_{out\,\omega''}a_{out\,\omega}\vert in\rangle\equiv\langle
out\,\omega\omega''\vert in\rangle =-e^{iW}\,(\alpha_{aa}^{-1}\beta_{ab}^*)
_{\omega\omega''}=\mp e^{iW}\,(\alpha_{bb}^{-1}\beta_{ba}^*)_{\omega''\omega},
\end{equation}                                                          
we note that it reduces to a product of the source amplitude $\beta_{ab}^*$ or
$\beta_{ba}^*$ of oppositely directed $a$ and $b$ particles and the
backscattering amplitude $\alpha_{aa}^{-1}$ or $\alpha_{bb}^{-1}$ of one of
them, as a result of which both particles of the pair move in the same direction.
The symmetry of Eqs.~(79) makes it impossible to establish which of the particles
of the $ab$ pair experiences backscattering.

\section*{Acknoledgements}
The author is grateful to A. I. Nikishov for stimulating discussions.  This work
was carried out with the financial support of the Russian Fund for Fundamental
Research (Grant No.~95-02-04219a).

\end{document}